\documentclass[prl,twocolumn,notitlepage,preprintnumbers,amsmath,amssymb,superscriptaddress]{revtex4-1}
\usepackage{graphicx,bm,amsmath}
\usepackage{amsmath,amssymb}
\usepackage{graphicx}
\usepackage{wasysym}
\usepackage{amsfonts}
\usepackage{bm}
\usepackage{enumerate}
\usepackage{color}
\usepackage{xcolor}
\usepackage{epstopdf}
\usepackage{latexsym}
\usepackage[breaklinks,colorlinks = true,linkcolor = red,urlcolor=cyan,citecolor=red]{hyperref}
\usepackage{stmaryrd} 
\usepackage{braket} 
\usepackage[textsize=tiny]{todonotes}

\newcommand{\br}{{\bf r}}

\newcommand{\bk}{{\bf k}}

\newcommand{\revise}[1]{#1}
\DeclareMathOperator{\Tr}{\mathrm{Tr}}
\DeclareMathOperator{\rIm}{\mathrm{Im}}

\usepackage[normalem]{ulem}

\usepackage{pdfpages}
\usepackage{pgffor}
\makeatletter
\AtBeginDocument{\let\LS@rot\@undefined}
\makeatother

\begin{document}
\title{Flat Topological Bands and Eigenstate Criticality in a Quasiperiodic Insulator}
\author{Yixing Fu}
\affiliation{Department of Physics and Astronomy, Center for Materials Theory, Rutgers University, Piscataway, NJ 08854 USA}
\author{Justin H. Wilson}
\affiliation{Department of Physics and Astronomy, Center for Materials Theory, Rutgers University, Piscataway, NJ 08854 USA}
\author{J. H. Pixley}
\affiliation{Department of Physics and Astronomy, Center for Materials Theory, Rutgers University, Piscataway, NJ 08854 USA}
\affiliation{Center for Computational Quantum Physics, Flatiron Institute, 162 5th Avenue, New York, NY 10010} 
 \affiliation{Physics Department, Princeton University, Princeton, New Jersey 08544, USA}
\date{\today}

\begin{abstract}
The effects of downfolding a Brillouin zone can open gaps and quench the kinetic energy by flattening bands. Quasiperiodic systems are extreme examples of this process, which leads to new phases and critical eigenstates. We analytically and numerically investigate these effects in a two-dimensional topological insulator with a quasiperiodic potential and discover a complex phase diagram. We study the nature of the resulting eigenstate quantum phase transitions; a quasiperiodic potential can make a trivial insulator topological and induce topological insulator-to-metal phase transitions through a unique universality class distinct from random systems.
This wealth of critical behavior occurs concomitantly with the quenching of the kinetic energy, resulting in flat topological bands that could serve as a platform to realize the fractional quantum Hall effect without a magnetic field.  
\end{abstract}

\maketitle

The interplay of topology and strong correlations produces fascinating phenomena, with the fractional quantum Hall effect \cite{stormer1999fractional} serving as the quintessential example.
Conventionally, the magnetic field induces topology in the electronic many-body wavefunction; however, Berry curvature of the band structure is sufficient to induce 
topological single-particle wavefunctions that
can survive  the presence of interactions (see Ref.~\onlinecite{Maciejko-2015} for a review). 
Despite strong numerical evidence of fractional Chern and $\mathbb{Z}_2$ insulators \cite{RegnaultBernevig2011FracChernIns,liu2012fractional,harper2014perturbative,bandres2016topological,maciejko2010fractional,swingle2011correlated}, 
identifying a clear experimental route to the many-body analog of the fractional quantum Hall effect without a magnetic field remains challenging.
\revise{Research in this direction has aimed to identify lattices with flat topological bands that quench the kinetic energy, promoting strong correlations 
\cite{bergholtz2013topological,parameswaran2013fractional,wang2012fractional,yang2012topological,heikkila2011flat,Lee-2016,Lee-2017}.}

Recent work on twisted graphene heterostructures opened up new platforms to study strongly correlated physics, including correlated insulators \cite{cao2018correlated}, superconductivity \cite{cao2018SC,yankowitz2019SC}, and Chern insulators \cite{sharpe2019emergent,pixley2019ferromagnetism,polshyn2020electrical}.
Proposals for realizing flat topological bands in these systems have followed  \cite{zhang2019flat,chittari2019gatetune,wu2019topological,wolf2018substrate, tong2017topological,san2014electronic, lian2019flat, ledwith2019fractional,song2019all}.
It was also recently shown in Refs.~\onlinecite{MASM, MASM_chiral} that the incommensurate effect of the twist could be emulated by a quasiperiodic potential.
Consequently, a class of models, dubbed magic-angle semimetals, show similar phenomena to twisted bilayer graphene (e.g., the formation of minibands and the vanishing Dirac cone velocity) at or near an eigenstate phase transition.
Similarly, to understand the theory for fractional Chern and $\mathbb{Z}_2$ insulators in incommensurate systems and how eigenstate criticality plays a role, it is essential to build a simple model to \revise{theoretically study} and \revise{experimentally realize}.
The notion of flat band engineering with incommensuration has broad applicability outside twisted heterostructures, including ultra-cold atomic gases \cite{kennedy2013spin, huang2016experimental, wu2016realization} and metamaterials~\cite{susstrunk2015observation,lustig2019photonic,peterson2018quantized,imhof2018topolectrical}.

\begin{figure}
    \includegraphics[width=0.9\columnwidth]{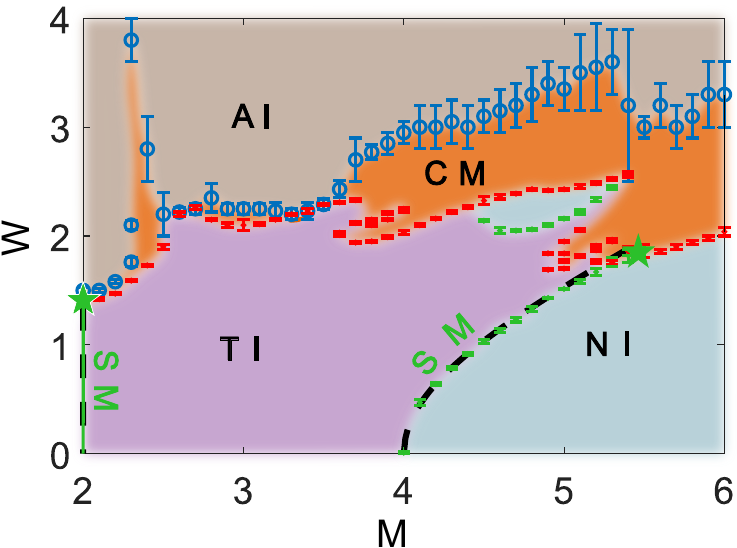}
    \caption{\revise{{\bf Phase Diagram}  of the 
    BHZ model in Eq.~\eqref{eqn:ham} at the band center
    with topological mass $M$ and quasiperiodic potential strength $W$.
    There are five illustrated phases: topological (TI), normal (NI), and Anderson (AI) insulators, Dirac semimetal (SM), and critical metal (CM).
    The green and red data points use the density of states in Eq.~\eqref{eqn:DOS} to locate the 
    transitions between TI and NI.
    Among them, the green data points and the green vertical line at $M=2$ are SMs, terminated at  magic-angle transitions (see \cite{SuppMat}) at the green stars.
    The black dashed lines are the perturbative prediction
    for the SM lines (e.g.  Eq.~\eqref{eqn:M}).
    The blue circles use transport [Eq.~\eqref{eqn:kubo}] to determine the   CM to AI boundary. 
   }
   }
    \label{fig:phase}
\end{figure}

In this letter, we study a minimal model for a two-dimensional topological insulator with a quasiperiodic potential to find a controllable route to create flat topological bands and induce quantum phase transitions beyond the Landau-Ginzburg paradigm, \revise{as there is no spontaneous symmetry breaking involved.
These transitions represent a universality class beyond the Altland-Zirnbauer classification of random matrices for disordered systems~\cite{Altland-1997,yamakage2013criticality}.
}
Using analytic and numeric techniques, we find an intricate phase diagram, as shown in Fig.~\ref{fig:phase}.
Particularly, quasiperiodicity creates practically flat topological bands near where finite-energy states exhibit criticality.
At the transition between topological and trivial insulators, the system realizes a magic-angle semimetal with features previously studied \cite{MASM}.
We further characterize the critical properties of the various eigenstate transitions, understanding them as localization and delocalization transitions in momentum- or real-space bases.
\revise{Importantly, these transitions and phases could be directly realized in twisted bilayer graphene that is close to aligned with a hexagonal boron nitride substrate~\cite{mao2020quasiperiodicity,PhysRevB.102.155136,PhysRevB.103.075122}.}

\revise{\emph{Model:}}
To describe a two-dimensional topological insulator, we use the Bernevig-Hughes-Zhang (BHZ) model~\cite{BHZ} with a 2D quasiperiodic potential.
The square-lattice Hamiltonian (with sites $\br$) is block diagonal 
\begin{equation}
  \mathcal H =  \sum_{\br,\br'} c^\dagger_{\br'} \begin{pmatrix} h_{\br'\br} & 0 \\ 0 & h_{\br'\br}^* \end{pmatrix}c_{\br}
  + \sum_\br c^\dagger_{\br} V(\br)c_{\br},
  \label{eqn:ham}
\end{equation}
where $c_\br$ are four-component annihilation operators and $V(\br) = W\sum_{\mu=x,y} \cos(Q r_\mu + \phi_\mu)$ is the quasiperiodic potential (QP) with amplitude $W$, wavevector $Q$, and random phase $\phi_\mu$; $h_{\br'\br}$ is a two-by-two matrix describing one block of the BHZ model ($h^*$, its complex conjugate). 
The nonzero elements of $h$ are \revise{$h_{\br\br} = (M-2t)\sigma_z$ and \revise{$h_{\br,\br+\hat{\mu}} = h_{\br,\br-\hat{\mu}}^\dagger = \tfrac12 t(-i \sigma_\mu +\sigma_z)$} for $\mu=x,y$ with Pauli matrices $\sigma_\mu$. Topological mass $M$ and the hopping $t=1$ set the energy scale.}
Most analyses are done on the two-by-two matrix since time-reversal symmetry relates each block, and $V(\br)$ does not couple blocks.
To reduce finite-size effects, we average over twisted boundary conditions implemented with $t \rightarrow t e^{i\tilde\theta_\mu/L}$ for a twist $\tilde\theta_\mu$ in the $\mu$-direction randomly sampled from $[0,2\pi)$.
The model is invariant under $M\rightarrow 4 - M$, so we focus on $M\geq 2$.
For $2<M<4$, the band structure (i.e., $W=0$) is topological with a quantized spin Hall effect $\mathcal{Q} = \sigma_{xy}^+ - \sigma_{xy}^-$ where $\sigma_{xy}^\pm$ are Hall conductivities for the blocks defined by $h$ and $h^*$ respectively. \revise{The superscript $\pm$ will be dropped as we focus on the $h$ block only.}
At $M = 2$ [$M=4$], the model is a Dirac semimetal with Dirac points at $\mathbf X = (\pi,0)$ and $\mathbf Y = (0,\pi)$ [$\mathbf M = (\pi,\pi)$] with velocity $v_0=t$.

Quasiperiodicity is encoded in $Q$, which in the thermodynamic limit we define as $Q/(2\pi) = (2/(\sqrt{5} + 1))^2$.
For simulations, we take rational approximates such that $Q\approx Q_L = 2\pi F_{n-2}/F_n$, where $F_n$ is the $n$th Fibonacci number, and the system size is $L = F_n$.
See the supplement for other values of $Q$.

\revise{\emph{Methods}: We investigate the phase diagram and phase transitions with spectral observables and eigenstates. 
Because the model in Eq.~\eqref{eqn:ham}
lacks translational symmetry, we 
treat the entire $L\times L$ system as a supercell, where the thermodynamic limit is
$L\rightarrow\infty$.
 At finite $L$, we define an effective band structure that is downfolded into a mini Brillouin zone (mBZ) of size $2\pi/L \times 2\pi/L$.}
\revise{
We apply the kernel polynomial method (KPM) \cite{weisse2006KPM} to compute spectral quantities and Lanczos or exact diagonalization to compute eigenstate properties (specifed in~\cite{SuppMat}). 
While the KPM and Lanczos work for larger $L$ than exact diagonalization, KPM introduces broadening to the data controlled by polynomial expansion cutoff $N_c$~\cite{weisse2006KPM} and Lanczos  limited to a small range of the spectrum.
}

\revise{To distinguish trivial, topological, and Anderson insulator phases, }we calculate
the conductivity tensor defined through Kubo formula \cite{garcia2015conductivity}
\begin{equation}
     \sigma_{\alpha\beta} = \frac{2 e^2 \hbar}{L^2}\! \int\! f(E)dE \rIm \Tr\left\llbracket v_\alpha  \frac{dG^-}{d\epsilon} v_\beta \delta(E - H)\right\rrbracket
     \label{eqn:kubo}
\end{equation}
where $f(E) = [e^{\beta(E-\mu)}+1]^{-1}$ is the Fermi function at inverse temperate $\beta$ and chemical potential $\mu$, $v_\alpha$ is the velocity operator,  $G^{-}$ is the retarded Green function, and $\llbracket \cdots \rrbracket$ denotes an average over phases in the QP ($\phi_{\mu}$) and twists ($\tilde{\theta}_{\mu}$) in the boundary condition. 
\revise{To determine phase boundaries and transition properties, we compute the density of states (DOS) which reflects band gaps and the low energy behavior of the semimetallic phase. 
The DOS at energy $E$ is }
\begin{equation}
    \rho(E) = \frac{1}{2L^2} \bigg\llbracket\sum_i \delta(E-E_i)\bigg\rrbracket
    \label{eqn:DOS}
\end{equation}
where $E_i$ denotes the energy eigenvalues.
\revise{
The gap centered at zero energy $\Delta$ is 
estimated with the KPM via the DOS satisfying
$\rho(E) < 0.001$ and with shift-invert Lanczos about $E=0$.
Along the semimetal lines the low-energy DOS goes like
$\rho(E)\sim \tilde{v}^{-2}|E|$, where $\tilde{v}$ is
the renormalized velocity of the Dirac cones that we
calculate through the scaling with $N_c$.
A detailed discussion of obtaining 
$\Delta$ and $\tilde{v}$
is in~\cite{SuppMat}.
}

For wavefunctions, we compute the inverse participation ratios (IPRs) in real and momentum space
\revise{
to discern localized, extended or critical states.
}
The IPR in a basis indexed by $\bm \alpha$ is 
\begin{equation}
    \mathcal{I}_{\alpha}(E) = \sum_{\bm\alpha} \left\llbracket\lvert\braket{\bm\alpha|\psi_E}\rvert^4\right\rrbracket
    \label{eqn:IPR}
\end{equation}
using normalized wave functions in the momentum space ($\bm{\alpha} = \bk$) or real space ($\bm{\alpha} = \br$) basis.
For systems localized in basis $\alpha$, $\mathcal{I_\alpha}$ is $L$-independent; for delocalized systems, it goes like $\mathcal{I}_{\alpha} \sim 1/L^2$. At a localization transition~\cite{Mirlin-2000,evers2008anderson} $\mathcal{I}_{\alpha} \sim 1/L^{\gamma}$ where $0<\gamma<2$ is the fractal dimension ($D_2$); \revise{$\gamma$ is extracted from the finite size effect when calculating $\mathcal{I}_\alpha$ at various system sizes~\cite{SuppMat}.}

\revise{To study band flatness and topology of the effective band-structure 
in the mBZ, 
we calculate the wavefunction $|\psi_{E_n}(\bm{\theta})\rangle$, where $\bm{\theta}$ is the crystal momentum of the $L\times L$ super-cell (via the twist in the boundary condition as $\bm \theta=\tilde{\bm{\theta}}/L$) and $E_n$ is the energy of the $n$th band labelled in ascending order. 
The bandwidth 
is then defined as 
$w_n=\max |E_{n}(\bm{\theta})-E_{n}(\bm{\theta}')|_{\bm{\theta},\bm{\theta}'}$
and the direct band gap is
$\Delta_{n}=E_{n+1}(\bm{\theta}) - E_{n}(\bm{\theta})$. 
The flatness ratio,
which measures a band's flatness and its isolation from the neighboring bands is defined following \cite{lee2016band} as
\begin{equation}
    f_n = \min\{\Delta_{n}, \Delta_{n-1}\} / w_n.
    \label{eqn:flatratio}
\end{equation}
The  Berry curvature $\Omega_n(\bm{\theta})$ and  Chern number $C_n$ can be determined via
momentum-space plaquettes
defined by the four momenta $\bm\theta\rightarrow\bm\theta_1\rightarrow\bm\theta_2\rightarrow\bm\theta_3\rightarrow\bm\theta$ 
\footnote{All plaquettes must be chosen with the same orientation.} 
and they can be calculated following \cite{fukui2005chern}
\begin{equation}
    \Omega_n(\bm{\theta}) = \rIm \ln 
    \frac{U_n(\bm{\theta},\bm{\theta}_1)U_n(\bm{\theta}_1,\bm{\theta}_2)}
    {U_n(\bm{\theta},\bm{\theta}_3)U_n(\bm{\theta}_3,\bm{\theta}_2)}, 
    \,
    C_n = \frac{1}{2\pi }\sum_{\bm\theta} \Omega_n(\bm{\theta})
    \label{eqn:berry}
\end{equation}
where $U_n(\bm{\theta}_a, \bm{\theta}_b)=\langle\psi_n(\bm\theta_a)|\psi_n(\bm\theta_b)\rangle/|\langle\psi_n(\bm\theta_a)|\psi_n(\bm\theta_b)\rangle|$
and the sum to obtain $C_n$ is over the mBZ $[0,2\pi/L)^2$.}
\revise{
 Lastly, we use machine learning of the wavefunctions 
 to provide
 an efficient survey of a large parameter space (in $W$, $M$, and $E$) as an additional validation of the phase diagram in Fig.~\ref{fig:phase}. This also reveals intriguing features of the Anderson insulating phase, as elaborated in~\cite{SuppMat}. }

\emph{Phase Diagram}:
\revise{Using diagrammatic perturbation theory and numerical calculations with the KPM and Lanczos}
we obtain the phase diagrams shown in Fig.~\ref{fig:phase}.
There are five phases pictured: topological insulator (TI), normal insulator (NI), critical metal (CM), Anderson insulator (AI), and lines of Dirac semimetals (SM) between TI and NI phases.
Both band-insulating and SM phases are stable to weak quasiperiodicity.
Finite band gaps and quantized (zero) spin Hall conductivity describe the TI (NI) phase, \revise{calculated using the KPM method with Eq.~\eqref{eqn:kubo}.
Low-energy scaling of the DOS $\rho(E)\sim \tilde v^{-2}|E|$ captures the SM phases (marked with green data points).
Other boundaries between gapped and finite DOS at $E=0$ are marked with red data points. These DOS results 
trace the phase boundaries between TI and NI (green) and between TI and CM (red).}
The AI phase has a finite DOS but zero conductivity and localized wave functions (i.e., real space IPR that is $L$-independent), \revise{with the phase boundary marked by blue circles with error bars}.
The structure revealed is $Q$-dependent~\cite{SuppMat} and reminiscent of other studies of insulating phases perturbed by quasiperiodicity \cite{roux2008quasiperiodic}.

\revise{Upon increasing $W$, for $M\lesssim4$ and $M\gtrsim5$ we traverse the phases TI/NI $\rightarrow$ CM $\rightarrow$ AI.
However, more complicated cuts are possible between $M=4.5$ and $M=5.3$, where 
quasiperiodicity drives trivial phases topological (for $4<M \lesssim 5.0$) and into-and-out-of metallic and topological phases at zero-energy.
An example is shown in the supplement \cite{SuppMat}, where increasing $W$ leads to the phases NI $\rightarrow$ SM $\rightarrow$ TI $\rightarrow$ CM $\rightarrow$ TI $\rightarrow$ SM $\rightarrow$ NI $\rightarrow$ CM $\rightarrow$ AI.}

The physics on the SM lines emanating from $M=2$ or $M=4$ at $W=0$ agrees with the universal features found in Ref.~\onlinecite{MASM} and reveals magic-angle transitions marked by green stars in Fig.~\ref{fig:phase}(a).
Concentrating on $M=2$, the semimetal is stable with a velocity (calculated from the DOS, \revise{see~\cite{SuppMat}}) that vanishes like $ \tilde v \sim (W_c(M=2)-W)^{\beta/2}$ where $W_c(M=2) = 1.42\pm0.02$ and $\beta=2\pm 0.3$, which is close to the universal value $\beta \approx 2$ obtained in other models and symmetry classes~\cite{MASM,MASM_chiral}.
\revise{
A CM phase is found as well as 
a localization transition at $W_A(M=2) = 1.50\pm 0.03$.
}

For smaller values of $W$, we use perturbation theory to map out the phase diagram and estimate the location of the NI-to-TI and SM-to-CM transitions \revise{(see~\cite{SuppMat})}.
\revise{These phase transitions can be assessed in this regime (i.e. near $M=4$) by computing the renormalized mass $\tilde{M}$ and velocity $\tilde v$.  We obtain up to second order in $W$ 
}
\begin{equation}
    \tilde M - 4 =   \frac{\left[(M-4) +W^2\frac{(4-M)+(\cos Q-1)}{(4-M)^2+2(3-M)(\cos Q-1)}\right]}{1+W^2/( (4-M)^2 + 2(3-M)(\cos Q - 1))}.
    \label{eqn:M}
\end{equation}
By solving for $\tilde M = 4$, we obtain the phase boundary between  insulating phases, illustrated by the black dotted line in Fig.~\ref{fig:phase}(a) (at fourth-order in $W$), which is in excellent agreement with the numerics. 
The curvature to this line demonstrates that quasiperiodicity can drive a topological phase transition NI-to-TI, which is the deterministic analog of the disordered topological Anderson insulator~\cite{groth2009theoryTAI,meier2018observation}.
For $M=2$, there is no renormalization of $\tilde M$.
Using numerics to access higher $M$ and $W$, when $M\gtrsim 5.4$, the NI transitions into the CM. The magic-angle transition (i.e., SM-to-CM) is obtained by solving $\tilde v \rightarrow 0$ on the line \revise{$\tilde M=4$. }

\begin{figure}[t!]
\includegraphics[width=0.95\columnwidth]{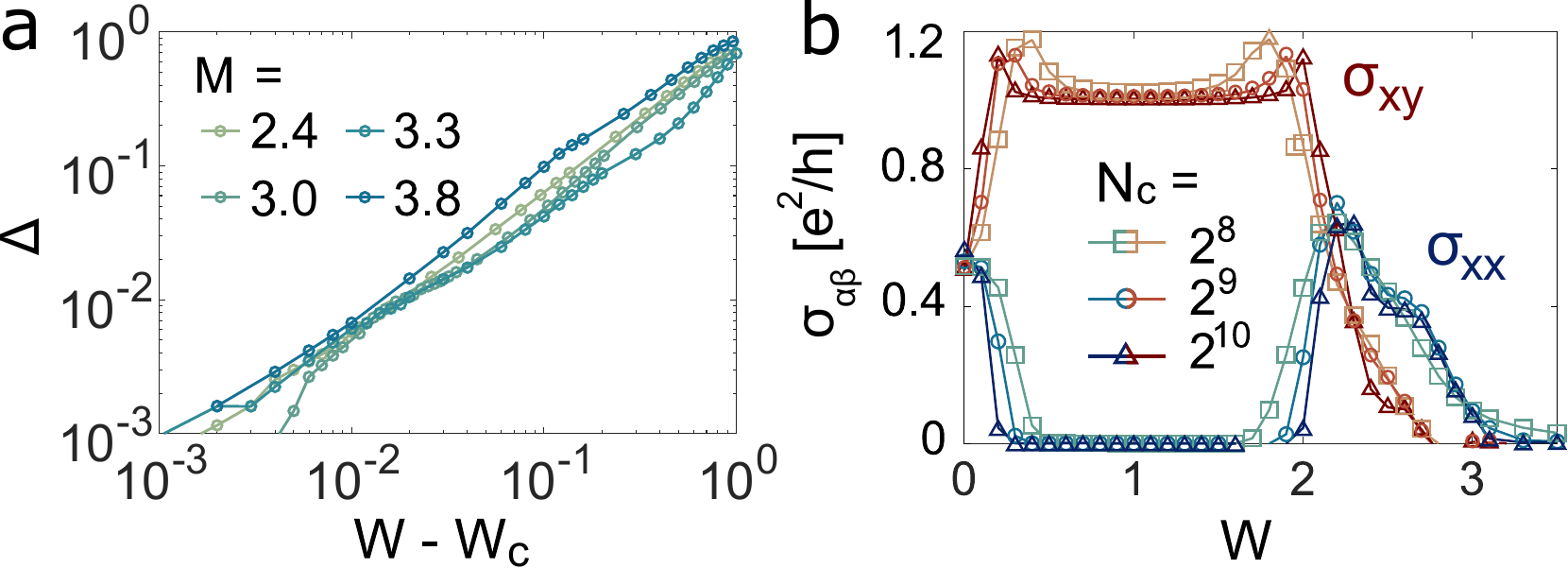}
\caption{
{\bf Demonstration of the TI-to-CM transition.}
(a) Tracking the density of states 
\revise{computed with the KPM in Eq.~\eqref{eqn:DOS}, we see the (hard) band gap closes 
as a power law} 
$\Delta = (W_c(M) - W)^{\nu z}$ and find $\nu z \approx 1$ at the TI-to-CM transition across each value of $M$. (b) Shows the conductivity \revise{computed with the KPM in Eq.~\eqref{eqn:kubo}} as a function of quasiperiodic strength $W$ for $M=4.0$. The Hall conductivity $\sigma_{xy}$ saturates to a finite value in the TI phase, but for \revise{$W_c(M=4)\approx 2\lesssim W \lesssim 3$} the longitudinal conductivity becomes finite 
and the Hall part is suppressed.
The system is localized when $W\gtrsim 3$. Note that the feature near $W=0$ is due to $M=4$ being a SM.
\revise{
We stress that this metallic phase and therefore this transition does not exist in the presence of randomness.
}
}
    \label{fig:TItoM}
\end{figure}

\emph{TI-to-CM transition}:
\revise{To analyze topological transitions that are forbidden in disorder systems we 
use numerics to capture the full, nonperturbative transition to the CM phase located at $W_c(M)$.
Near the transition, the correlation length diverges as $\xi\sim |W-W_c|^{-\nu}$ while scale invariance implies that the gap $\Delta\sim \xi^{-z}$; therefore the gap vanishes like $\Delta\sim|W-W_c|^{\nu z}$.}
\revise{Through the KPM calculation of DOS and Lanczos calculation of lowest energy states, we find $\nu z\approx 1$ for each $M$ value we have considered, see Fig.~\ref{fig:TItoM}(a).}

\revise{These exponents indicate a unique universality class driven by quasiperiodicity distinct from random systems.
Since our system breaks up into two blocks, each in the same symmetry class as the quantum Hall effect (i.e., class A), random disorder does not allow for a metallic phase~\cite{evers2008anderson,yamakage2013criticality,chen2015tunable}. 
Therefore, topological phase transitions driven by quasiperiodicity host unique universality classes beyond the ten
Altland-Zirnbauer random matrix classes~\cite{Altland-1997}.
}

When gap closes at $W_c(M)$, the conductivity at $E=0$ becomes finite, and the Hall conductivity is no longer quantized, indicating the onset of the CM phase.
As seen in Fig.~\ref{fig:TItoM}(b), the Hall conductivity drops, and $\sigma_{xx}$ peaks at the transition, remaining finite for the duration of the CM.
\revise{
The transition does not involve any symmetry breaking; it occurs when the topological gap closes and $\sigma_{xy}$ is no longer quantized.}
For larger values of $W$, we find a transition into an Anderson insulating phase \cite{evers2008anderson, abrahams1979scaling} with exponentially localized wavefunctions in real space and a vanishing $\sigma_{xx}$.

\begin{figure}[t!]
\includegraphics[width=0.96\columnwidth]{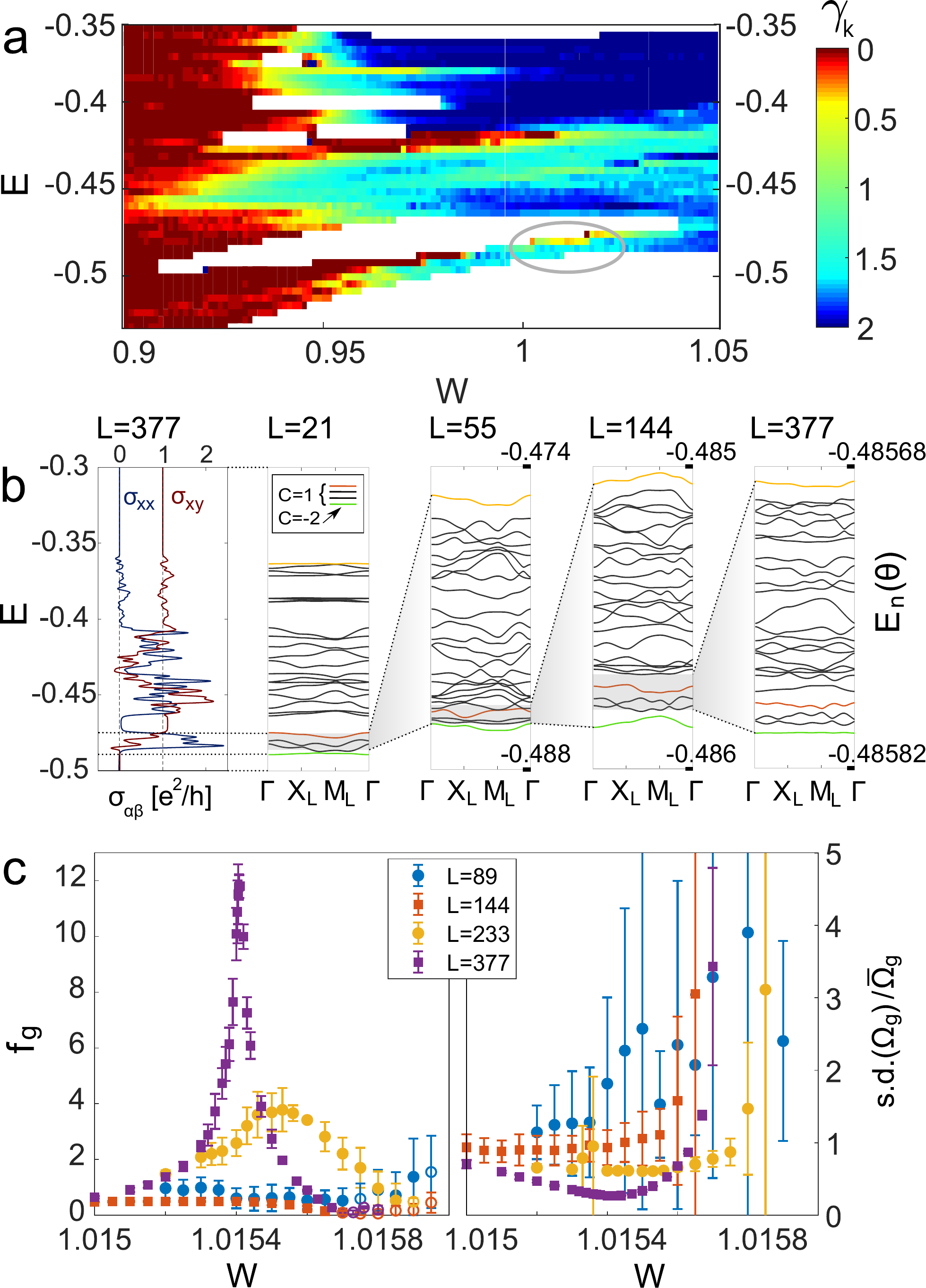}
    \caption{{\bf Flat Chern bands and eigenstate criticality.} 
    \revise{(a) Color plot of the momentum-space IPR system-size scaling (as defined in Eq.~\eqref{eqn:IPR}). 
    Notice that around $W\sim 0.95$ low energies become delocalized in momentum space while at higher energies $\mathcal{I}_k \sim L^{-\gamma_k} $ for $0<\gamma_k<2$ indicating critical eigenstates along the mobility edge; the value of $\gamma_k$ is given by the color.
    The lowest energy states (and narrowest set of states) has a Chern number of 1. The white regions are hard gaps.
    (b; left) the conductivity calculated from Eq.~\ref{eqn:kubo} with $L=377$ and cutoff $N_c=2^{14}$. (b; right) Dispersion relation $E_n(\bm{\theta})$ along a representative cut in the mBZ 
    for a sequence of $L=F_{n}$ with even $n$, for $W=1.0154$.
    For each $L$, the green band carries Chern number $-2$, the first 4 bands (from green to cyan) sum to Chern number $1$, and the 25 bands pictured in each plot sum to Chern number $1$ (for $L=55$, the pattern appears to hold but the lowest bands do not have a well-defined gap).
    (c) the flatness ratio $f_g$ (left, as defined in Eq.~\ref{eqn:flatratio}) and
     the normalized standard deviation of Berry curvature $\Omega_g$ (as defined in Eq.~\ref{eqn:berry}) across the folded Brillouin zone (right) of the first band above $E=-0.5$, for various $L$ values. 
     For $L=233$ and $L=377$, the peak of the flatness ratio appears near where the Berry curvature has less fluctuations. 
     The filled markers ($\bullet$) indicate topological bands while empty markers ($\circ$) indicate trivial bands (excluded in the right). The squares ($\blacksquare$) and circles ($\bullet$) correspond to $L=F_n$ such that $n$ is odd and even, respectively. 
    }
    }
    \label{fig:flatChern}
\end{figure}

\emph{Criticality and flat topological bands}:
\revise{At small $W$, the insulating band gap [computed via the DOS  in Eq.~\eqref{eqn:DOS}] increases for some values of $M$ but decreases for larger $W$, 
which is beyond the
 perturbative theory in Eq.~\eqref{eqn:M}.  
This non-monotonicity is demonstrated in  \cite{SuppMat} and coincides with the onset of criticality in the finite energy 
states (i.e. a mobility edge) near
the edge of the gap centered about $E=0$ (e.g. in Fig.~\ref{fig:flatChern} this corresonds to the states near $ E \approx -0.5$ for $W\approx 1$).
Due to the interplay of topology, criticality, and quasiperiodicity
several physically interesting effects occur near the gap maximum. This is demonstrated in Fig.~\ref{fig:flatChern} for $M=4.0$ as a representative cut of the phase diagram in Fig.~\ref{fig:phase} that we now explore in more detail.
}

\revise{It can be seen from Fig.~\ref{fig:flatChern}(a) that the states 
\footnote{By downfolding this collection of states originated from a band at a given $L$} 
near $E\approx-0.5$ narrow around $W\approx1$ and are well isolated from other states by hard gaps. 
Additionally, by calculating $\sigma_{xy}$, Fig.~\ref{fig:flatChern}(b) (left most panel) shows this collection of bands has total Chern number equal to $1$, independent of $L$. 
Meanwhile, these states become critical, as measured by the IPR in momentum and position space ($1/\mathcal{I}_{\alpha}\approx L^{\gamma_{\alpha}}$) with $0<\gamma_{\alpha}<2$, showing they are delocalized in both bases ($\alpha=x,k$) [Fig.~\ref{fig:flatChern}(a) where color shows $\gamma_k$].
}
\revise{Interestingly, we observe a
self-similarity in these critical bands;
the sequence of decreasing energy windows shown in Fig.~\ref{fig:flatChern}(b) have the same Chern number as we increase the super-cell size.
When $M=4$ and $W=1.01541$ the relevant energy window $E\in [-0.49,-0.47]$  has $(F_{n-5})^2$ states for a system size $L=F_n$.
When we can identify the lowest band [depicted by the green line in Fig.~\ref{fig:flatChern}(b)] in this energy window
the value of its Chern number 
follows the self similar sequence 
of $C=-2$ for $L=F_{2n}$ and $C=1$ for $L=F_{2n-1}$
(in each case examined). 
}

\revise{
The flatness of the lowest (green) band is 
apparent in the dispersion in the mBZ  in Fig.~\ref{fig:flatChern}(b) as well as 
by its large effective mass~\cite{SuppMat}.
By computing the flatness ratio (of the green band) $f_g$ and Berry curvature $\Omega_g$ (in Eqs.~\eqref{eqn:flatratio} and \eqref{eqn:berry}, respectively) our data also demonstrates that larger $L$ leads to flatter, isolated topological bands in the mBZ at some optimal $W$. As shown in Fig.~\ref{fig:flatChern}(b and c left)  for increasing $L$ the peak in $f_g$ sharpens  concomitantly with the  development of critical eigenstates [Fig.~\ref{fig:flatChern}(a)]
as the Chern bands in the mBZ occur at an increasingly fine energy scale. 
At the $W$ with maximal $f_g$, we also see 
a reduction in the fluctuation in Berry curvature $\Omega_g$ (of the green band), probed via its standard deviation divided by the mean across the mBZ~\cite{SuppMat}, Fig.~\ref{fig:flatChern}(c,right). 
The reduction of fluctuations of 
$\Omega_g$
for increasing $L$ suggest that the model could host a fractional Chern insulating state in the presence of interactions~\cite{parameswaran2012fractional,Claassen-2015}; however, it is possible that 
an incommensurate charge density wave state could be stabilized instead (see~\cite{SuppMat} for Berry curvature profiles in the mBZ).
}

\emph{Conclusion--}
\revise{In a simple model of a two-dimensional topological insulator, we demonstrated that the inclusion of quasiperiodicity induces flat bands, eigenstate criticality, and a phase diagram full of structure.}
The eigenstates go through several Anderson-like transitions (delocalizing in momentum space before localizing in real space), which leads to critical eigenstates in a metallic phase.
Meanwhile, we see the onset of flat topological bands within the TI phase concomitant with critical high energy eigenstates.
\revise{Our results go beyond twisted heterostructures
and} allows for cold atom labs and metamaterial labs (both of which have already realized 2D TIs 
\cite{peterson2018quantized,imhof2018topolectrical,susstrunk2015observation,lustig2019photonic,kennedy2013spin,huang2016experimental, wu2016realization}) to emulate similar physics.

\begin{acknowledgments}
We thank Yafis Barlas, Elio K\"onig, and Jie Wang for useful discussions.
This work is partially supported by  Grant No.\ 2018058 from the United States-Israel Binational Science Foundation (BSF),  NSF CAREER Grant No. DMR-1941569, and  by the Air Force Office of Scientific Research under Grant No.~FA9550-20-1-0136.
Numerical calculations were done using Julia \cite{bezanson2017julia}.
The authors acknowledge the following research computing resources that have contributed to the results reported here: 
the Open Science Grid~\cite{osg07,osg09}, which is supported by the National Science Foundation award 1148698, and the U.S.\ Department of Energy's Office of Science,
the Beowulf cluster at the Department of Physics and Astronomy of Rutgers University; and the Office of Advanced Research Computing (OARC) at Rutgers, The State University of New Jersey (http://oarc.rutgers.edu), for providing access to the Amarel cluster.
The Flatiron Institute is a division of the Simons Foundation.
\end{acknowledgments}

\bibliography{QP_BHZ}



\section*{Supplementary material}

# Supplement to "Flat Topological Bands and Eigenstate Criticality in a Quasiperiodic Insulator"

Yixing Fu,[1] Justin H. Wilson,[1] and J. H. Pixley[1, 2, 3]

[1]*Department of Physics and Astronomy, Center for Materials Theory, Rutgers University, Piscataway, NJ 08854 USA*
[2]*Center for Computational Quantum Physics, Flatiron Institute, 162 5th Avenue, New York, NY 10010*
[3]*Physics Department, Princeton University, Princeton, New Jersey 08544, USA*
(Dated: April 11, 2021)

In the following supplemental material we provide details about the phase diagram, the magic-angle transition along the semimetal lines, the machine learning algorithm we have used to support the phase diagram in text, the perturbation theory at second and fourth order in the potential, as well as additional numerical results.

## CONTENTS

## I. ADDITIONAL PROPERTIES OF THE PHASE DIAGRAM

### A. Main Phase Diagram

Here we augment the phase diagram shown in Fig. 1 of main with several details. Focusing on Fig. S1(left), the magenta line indicates an additional measure of the Anderson localization transition, using a neural net model as elaborated in Section III. Critical eigenstates may appear localized and therefore are not straightforward to diagnose with the IPR and conductivity alone (due to their anomalous scaling with system size). Thus, we use machine learning to provide a conservative measure of the Anderson localized phase, which in certain regimes matches the IPR and conductivity, but in other regimes of the phase diagram extends to larger values of $W$. In addition, the machine learning approach identifies an area ranging from $M = 4$, $W = 3$ to $M = 4.5$, $W = 4$ in the phase diagram (the darker region bounded by pink curve in Fig. S1) as critical, that is not identified by other observables. We leave the details on the CM-AI phase transition for future studies.

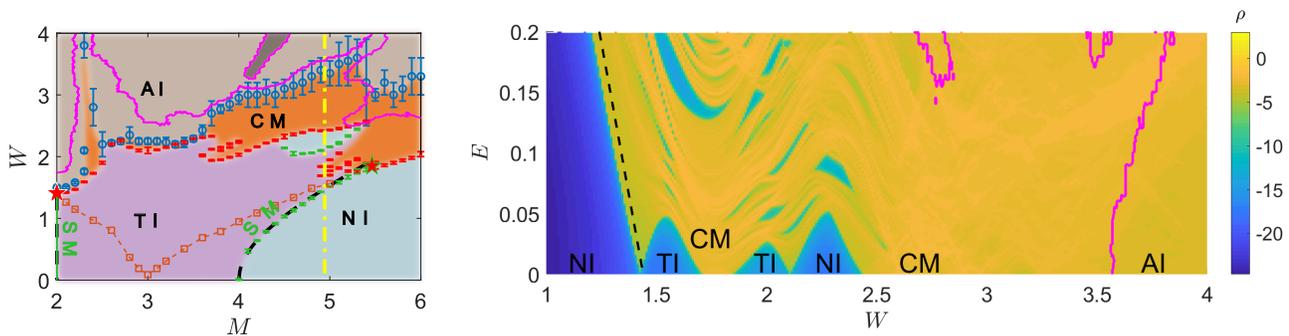

FIG. S1. (a) Full phase diagram with all measures used to diagonose phases and transitions. The magenta line shows the boundary between delocalized or critical phase and localized phase at zero energy, as indicated by the neural net model. The dark region, roughly extending from $M = 4$, $W = 3$ to $M = 4.5$, $W = 4$ is indicated as critical phase by the neural net model, but not identified by any other observables. The dashed orange line inside the TI phase shows where the size of the gap centered at $E = 0$ is maximals and thus starts to significantly deviate from perturbation theory. (b) A cut of the phase diagram in energy space represented by the yellow line in (a). Notice the multiple phase transitions, all driven by quasiperiodicity ($W$) and the higher energy metallic nature. The pink curve represents the boundary to machine-learned, localized eigenstates.

The yellow line starting from $M = 4.9$ is expanded in the parameter space of $W, E$ in Fig. S1(right). The cut show an example of the very rich sequence of phase transitions, into-and-out-of metallic and topological phases at zero energy, made possible by only increasing $W$ at fixed $M$. The magenta curve show the boundary to localized states as determined by the neural net model.

Inside the TI phase with intermediate $W$, IPR of lowest energy states demonstrate a region of critical scaling well before entering CM phase. Such region of criticality can provide a coarse guide to the parameters $(M, W)$ where flat topological band discussed in the main text is developed. Intriguingly, the onset of such criticality can roughly be traced by the dashed orange line (with □ symbols) originating from $M = 3$, $W = 0$ in Fig. S1(a) where the size of the gap centered at $E = 0$ no longer follows perturbation theory (see Section IV for perturbation theory and Section VI for numerical calculation of the gaps). The eigenstates at the band edge appear critical, as measured by the IPRs, for a range of $W$ above such line (see Section VI).

## B. Dependence on the quasiperiodic wavevector $Q$

In the main text we have focused on a quasiperiodic wavevector $Q = 2\pi F_{n-2}/F_n$ and linear system size $L = F_n$, where $F_n$ is the $n$th Fibonnaci number. The structure of the phase diagram is strongly dependent on the incommensurate value of $Q$ chosen. Here, we show the zero energy density of states that probes insulating, semimetallic, and metallic parts of the phase diagram (but cannot discern between delocalized and localized wavefunctions) in Fig. S2 for $Q = F_{n-3}/F_n$ (a) and $Q = F_{n-4}/F_n$ (b). With these smaller values of $Q$, the semimetal-to-metal magic-angle transition along $M = 2$ happens for smaller values of $W$. Qualitatively, this behavior is captured by the perturbation theory near $W = 0$, where it shows the Dirac cone velocity $v$ is renormalized more strongly for smaller values of $Q$. However, for even smaller $Q$, higher order perturbation theory is required to see the velocity renormalizing down to 0. On the other hand, the phase boundary rooted from $M = 4$ and $W = 0$ along the NI-to-TI phase boundary can be predicted well by the perturbation theory (shown as a red line in Fig. S2).

## C. Chemical Potential dependence of $\sigma_{xy}$

The Chern number is directly determined by the Hall conductivity $\sigma_{xy}$. While we use $\rho(0)$ to accurately locate phase boundaries, the numerical calculation of $\sigma_{xy}$ at large system sizes has more computational complexity. However, $\sigma_{xy}$ can be used to distinguish trivial and topological states and locate where the Chern bands are in energy. Here, we show an example of a color plot of $\sigma_{xy}$ at a fixed $M = 4.2$ and varying the Fermi energy $E_F$ and disorder strength $W$, see Fig. S3. The emergence of a topological phase after the collapse of the (lowest energy) trivial band gap can be seen clearly. In addition, our perturbative result (shown as the red lines) is in excellent agreement with the numerics in locating these phase boundaries for $0 \le W \lesssim 2$.

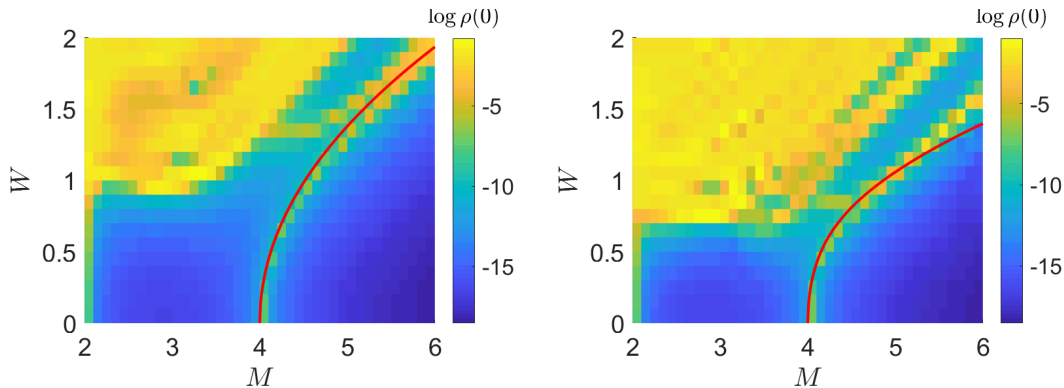

FIG. S2. **Phase diagram determined from density of states at** $E_F = 0$ **for** $Q = F_{n-3}/F_n$ **and** $Q = F_{n-4}/F_n$. The red line shown is the result of fourth order perturbation theory used to determine the location of the NI-to-TI transition from the vanishing of the renormalized topological mass $\tilde{M}$. The perturbative result agrees well with the numerics up to $W \approx 1$.

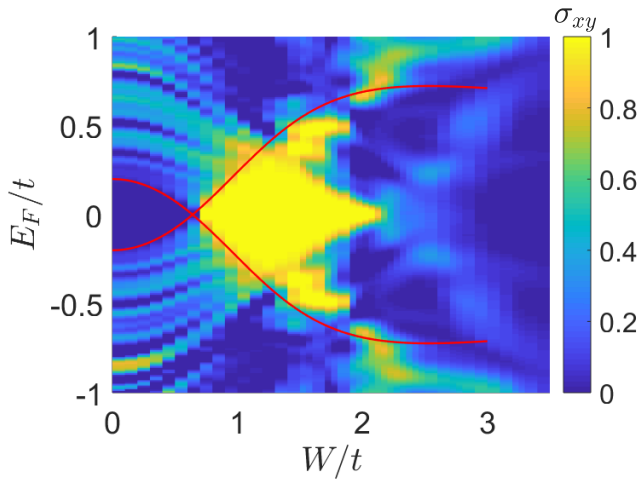

FIG. S3. **Finite energy topological phase diagram**. The Hall conductivity $\sigma_{xy}$ at various Fermi energies $E_F$ and quasiperiodicity $W$. The red lines are the perturbation theory prediction of gap size.

## II. MAGIC-ANGLE TRANSITION

The semimetal at $M = 2$ is stable with a velocity (calculated from the DOS) that vanishes like $\tilde{v} \sim (W_c(M = 2) - W)^{\beta/2}$ where $W_c(M = 2) = 1.42 \pm 0.02$ and $\beta = 2 \pm 0.3$. This is demonstrated in Fig. S4(a) where $\tilde{v}$ vanishes when $\rho(0)$ rises. Additionally, the wave functions are localized in momentum space when $W < W_c(M = 2)$, and delocalized in momentum space when $W > W_c(M = 2)$ (as indicated in Fig. S4(b) by $\mathcal{I}_k$ being $L$-independent and $\mathcal{I}_k \sim 1/L^2$, respectively). When the real space IPR is $L$-independent and the resistivity increases with $L$ and $N_c$, there is a localization transition with $W_A(M = 2) = 1.50 \pm 0.03$, indicating a small but finite CM phase.

## III. MACHINE LEARNING THE LOCALIZATION TRANSITION

In the present model, we found it challenging to pinpoint the Anderson localization transition using conductivity and the inverse participation ratio due to a large number of critical states that can appear localized by some metrics but not others. Therefore, we have supplemented this analysis with a machine learning classification of the single particle wavefunctions.

Machine learning is a class of methods where a non-specialized program can be used to perform a specific task when supplied with an abundance of data. Many machine learning techniques have been applied to various aspects of physics[S1–S3]. In this work we used Convolutionary Neural Networks (CNN) to distinguish whether a wave function is localized or extended. We train the neural network on a set of wave functions whose localized nature can be easily and

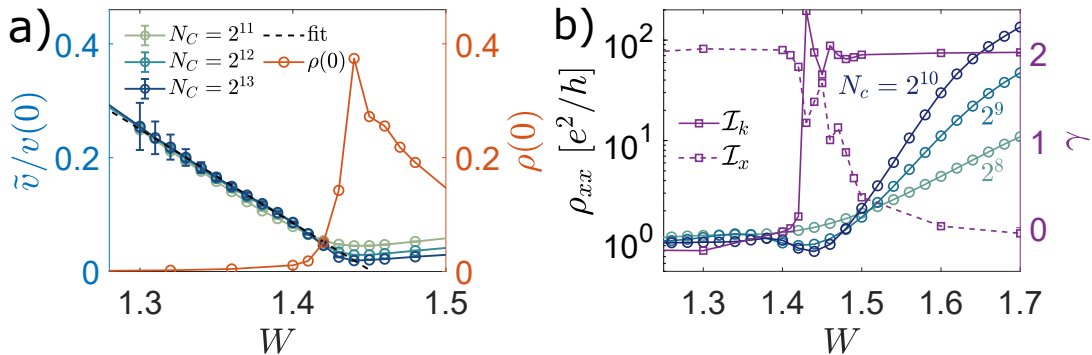

FIG. S4. **The magic-angle transition for the semimetal line** $M = 2$. (a) Renormalized velocity $v/v(0)$ and the resulting finite density of states $\rho(0)$ at the transition, extracted from $\rho(E)$ that is calculated using KPM method with system size $L = 144$, Chebyshev cutoff $N_C = 2^{15}$. See Fig. S7 for further discussion on obtaining $\tilde{v}$. (b) These plots indicate the appearance of a critical metallic phase $1.4 \lesssim W \lesssim 1.5$ inferred from both the resistivity $\rho_{xx}$ and the scaling of the momentum- and real-space IPRs. $\rho_{xx}$ is calculated using Kubo formula (see main text) with KPM method. The $L$-dependence of the IPRs is fitted from lowest energy eigenstates obtained using Lanczos method for $L = 89$, $L = 144$, and $L = 233$ to a power law form $\mathcal{I}_\alpha \sim 1/L^{\gamma_\alpha}$, and $\gamma_\alpha$ is shown as the right vertical axis.

unambiguously determined. Such a neural network automatically applies to all other points in the phase diagram, determining the phase boundaries efficiently and objectively.

A neural network consists of a massive number of nonlinear functions and linear transformations, usually as several "layers," to replicate any task that distills information from data. Practically, such a combination can be tuned to fit any mapping. Hence, as long as a concrete definition of the task to be executed is available, we can use labelled data as an example to tune the neural network until it replicates the task. Such a process is called "training," and can be calculated efficiently using modern computers.

In the present context, the problem that we want the neural network to solve is to distinguish localized wavefunctions from extended ones. This task can be thought of as a mapping from the space of all wavefunctions to a binary result of localized or extended. Using a set of wavefunctions labelled in advance, we can train the neural network to capture the relation between wavefunction data and the prediction of a localized phase. Once the training is finished, we can use the neural network model to classify a much larger dataset of wavefunctions, and map out a detailed phase diagram.

One crucial but more technical component of deep learning is the choice of the form and organization (i.e., "architecture") of the nonlinear functions used to fit[S4,S5]. In this work, we use a simple version of CNN. The wavefunction classification task is somewhat analogous to figuring out whether the image includes a dog or cat, which is a classical application of CNN. The CNN architecture makes use of a convolution operation prior to applying the nonlinear functions. The convolution operations effectively scrambles but preserves the information at various locations of the input data, and hence makes the model "translational invariant", i.e. the location of the feature does not affect the output. Such translational invariance allows the neural net model to treat critical and/or localized structures at different locations in the same way.

The neural network methods of machine learning usually suffer from over-fitting that harms the predictive power of the model. Simple and conventional methods against over-fitting including adding regularization terms, use of drop-off layers[S6] and so on. These methods are practically efficient and sufficient for our purpose.

A summary of the architecture we have used with a convolutionary neural network and drop-off layers is shown in Fig. S5. We experimented with a few different hyper parameters of the CNN architectures, and the model yields similar results. Hence the neural net model we have used is not a result of fine tuning. The robustness across different setups is likely because the localization feature is prominent and less ambiguous as opposed to typical computer vision tasks.

The training set is constructed in two different ways:

1. We look at the images to judge whether each wavefunction is clearly localized or not. The cases in which we are unsure are discarded from the training set. To minimize the affect of systematic bias caused in labelling the training set, we go through wavefunctions at several runs where each set is drawn randomly from the entire collection of wavefunctions and shuffled. Hence, the mislabelling can be considered as a random variation that is independent from the features that do not affect decision boundaries.

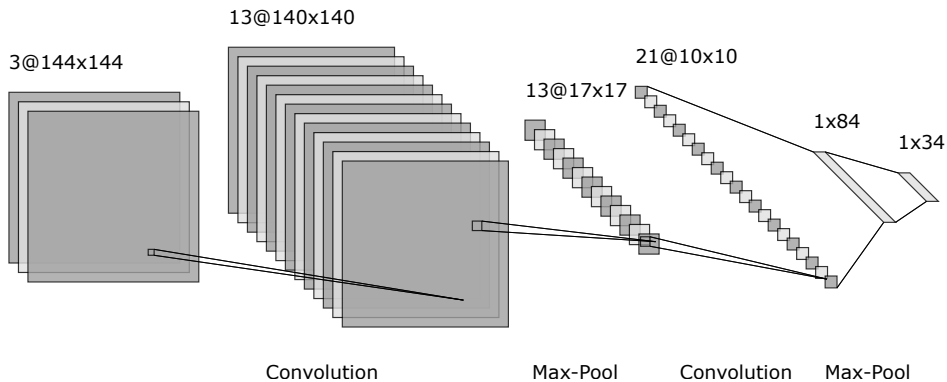

FIG. S5. **Schematic diagram of the neural network structure used for localization detection**. For convolution layers, we apply a convolution operation over a small window to get a data point in the next layer. Max-Pool layer simply takes the maximum of each window to reduce the model size. We also add batch-normalization and dropout layers before and after Max-Pool, but they are not shown here as they do not alter the overall architecture.

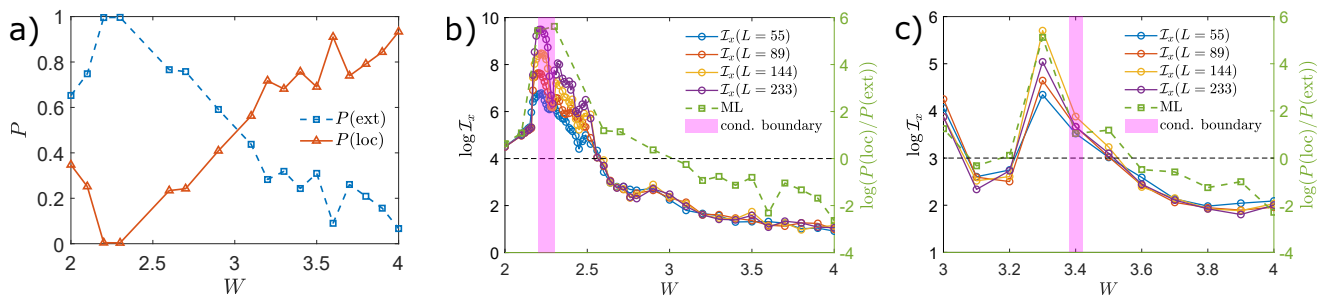

FIG. S6. **Comparing the IPR with the machine learning outcome**. (a) Shows an example of the neural network output for $M = 2.7$, given as the probability of a state being localized [$P(\text{loc})$] or extended [$P(\text{ext})$]. The summarized results are shown for $M = 2.7$ (b) and $M = 4.9$ (c), with comparison against KPM and IPR results. The difference between the two probabilities measures how confidently the model can distinguish localized or extended. Also shown in the figure with the magenta strips is the phase boundary determined by the conductivity, which indicates a transition near $W = 2.25$ for $M = 2.7$, and $W = 3.4$ for $M = 4.9$. Although the three different methods match quite well for $M = 4.9$, for $M = 2.7$ the IPR shows strongly critical behavior up until $W = 2.5$, well after the conductivity appears to vanish. Such critical behavior is detected by the neural net model. For $W$ between 2.3 and 2.5 the IPR shows a strong $L$ dependence and the neural net model predicts an extended phase with high confidence. For a range of $W$ larger than 2.5, the IPR shows a weak $L$ dependence across different system sizes, while in the neural net model $P(\text{loc})$ and $P(\text{ext})$ are quite close to each other.

2. We choose $W > 6$ for localized wavefunction examples, and sample $W = 0$ at various values of $M$ for extended wavefunctions.

The training set of method 2 does not include any of the critical wavefunctions in the CM phase. As a result, the CNN model identifies the critical phase as localized, producing a phase boundary in line with SM/TI to CM transition. This result can also be replicated using the training set from method 1 if we only include extended and fully localized wavefunctions. However, with method 1 we can instead label a dataset such that the non-localized label *includes* critical wavefunctions to provide an interesting complement to the KPM results and is hence included in the main results of Figure 1.

The phase boundary obtained from machine learning between localized and non-localized wavefunctions roughly traces the CM-AI phase boundary provided by the conductivity computed with the KPM for $M$ between 3.8 and 5.4, but it provides a slightly different boundary elsewhere. For $M < 3.8$, the machine learning result labels regions as critical that have a conductivity that looks localized (i.e. $\sigma_{xx}$ is vanishing with increasing $N_c$). We further investigate the nature of this region using the inverse participation ratio (IPR) in real and momentum space bases, see Fig. S6. The IPR in this region shows critical behavior that transits into a localized phase at a point that is hard to accurately determine. The machine learning result provides a conservative estimate of where the criticality ends and localization sets in.

In summary, our use of the machine learning method in the present context is to provide an additional measure of

the non-trivial phase boundaries that have a lot of structure. We then use conventional methods (conductivity and the IPR) to validate the physical nature of the phase boundaries.

## IV. PERTURBATION THEORY AT SECOND AND FOURTH ORDER

In this section, we provide additional details on the perturbation theory at second and fourth order that correctly captures the renormalization of the topological mass (to describe phase transitions in and out of the TI) and the velocity in the SM phase.

We begin by considering the single-particle Green function

$$\hat{G}_0(\omega) = [\omega - h_0(\omega)]^{-1}, \quad \hat{G}(\omega) = [\omega - h_0(\omega) + V]^{-1} \tag{S1}$$

and use Dyson's equation

$$G(\mathbf{k}, \omega)^{-1} = \omega - h_0(\mathbf{k}) - \Sigma(\mathbf{k}, \omega) \tag{S2}$$

where $\Sigma(\mathbf{k}, \omega)$ is the self-energy at momentum $\mathbf{k}$ including all $G_0(\mathbf{k}, \omega)$ irreducible diagrams. Close to the SM phase near $M = M_1 \equiv 2$ or $M = M_2 \equiv 4$, we express the Hamiltonian in the low-energy limit around the corresponding Dirac node $\mathbf{K}$ as $h_0(\mathbf{K}+\mathbf{q}) = v\mathbf{q}\cdot\boldsymbol{\sigma} + (M-M_i)\sigma_z$ and similarly expand the self-energy to obtain $\Sigma(\mathbf{k}=\mathbf{K}+\mathbf{q},\omega) = \omega\Sigma_E\sigma_0 + \Sigma_p\mathbf{q}\cdot\boldsymbol{\sigma} + \Sigma_z\sigma_z$ (where $\sigma_0$ is the 2-by-2 identity matrix and the $\sigma_{x,y,z}$ are the Pauli matrices). We define the quasiparticle residue $Z$, the renormalized topological mass $\tilde{M}$. and renormalized velocity $\tilde{v}$ such that the resulting Green function in the low-energy limit has the form

$$G(\mathbf{k} = \mathbf{K} + \mathbf{q}, \omega) = \frac{Z}{\omega - \tilde{v}\mathbf{q}\cdot\boldsymbol{\sigma} - \tilde{M}\sigma_z}. \tag{S3}$$

Then using $\Sigma_E$, $\Sigma_z$ and $\Sigma_p$ from $\Sigma(\mathbf{k}, \omega)$ we can express $Z$, $\tilde{M}$ and $\tilde{v}$ as:

$$Z^{-1} = 1 - \Sigma_E, \tag{S4}$$

$$\tilde{M} - M_i = (M - M_i + \Sigma_z)Z^{-1}, \tag{S5}$$

$$\tilde{v} = v(1 + \Sigma_p/v)Z^{-1}, \tag{S6}$$

To calculate $\Sigma(\mathbf{k}, \omega)$, we treat $V(\mathbf{r})$ perturbatively. In momentum space, $V$ is a delta function connecting $\mathbf{k}$ to $\mathbf{k} \pm Q\hat{x}$ and $\mathbf{k} \pm Q\hat{y}$. Hence, at second order the self energy is

$$\Sigma^{(2)}(\mathbf{k}, \omega) = (W/2)^2 \sum_{\pm, \hat{\mu}=\{\hat{x}, \hat{y}\}} \frac{1}{\omega - h_0(\mathbf{k} \pm Q\hat{\mu})}. \tag{S7}$$

Near $M = 4$, $\mathbf{k} = \mathbf{M} + \mathbf{q}$ with $\mathbf{M} = (\pi, \pi)$, this yields

$$\Sigma_E^{(2)} = -\frac{W^2}{D_2}, \tag{S8}$$

$$\Sigma_p^{(2)} = \frac{W^2}{2} \frac{(4-M)^2(1+\cos Q)}{D_2^2} v, \tag{S9}$$

$$\Sigma_z^{(2)} = W^2 \frac{(4-M) + (\cos Q - 1)}{D_2}, \tag{S10}$$

where $D_2 = (4-M)^2 + 2(3-M)(\cos Q - 1)$ is the common denominator that is always positive for $M > 3$. Observe that the numerator of $\Sigma_z^{(2)}$ is also always negative for $M > 4$, and $\Sigma_E^{(2)}$ is always negative. Hence $\tilde{M}$ is renormalized to be smaller as $W$ increases, predicting a critical $W$ where $\tilde{M}(W) = 4$ where TI to CM transition occurs. On the other hand, the direction of velocity renormalization is not obvious from the second-order perturbation theory, and indeed we can only predict the velocity to be renormalized to 0 at fourth-order perturbation theory. This is indicative of scattering off a single Dirac cone, where due to spin selection rules it requires a larger momentum exchange to induce intranode scattering.

The fourth-order perturbation theory includes all of the diagrams that connect the Dirac node to points in the Brillouin zone that are $2Q$ Manhattan distance apart and then back. The fourth order contributions to $\Sigma(\mathbf{k}, \omega)$ are

$$\Sigma_E^{(4)} = \frac{W^4}{4}(-15M^4 + 166M^3 + (-36M^2 + 206M - 295)\cos(3Q) - 732M^2 +$$
$$(2M(M(24M - 221) + 697) - 1497)\cos(Q) + (2M(M(13M - 115) + 356) - 770)\cos(2Q) +$$
$$6(M - 3)\cos(4Q) + 1522M - 1260)/D_4 \quad \text{(S11)}$$

$$\Sigma_z^{(4)} = \frac{W^4}{8}(-10M^5 + 138M^4 - 806M^3 + 2509M^2 + (2M(2M(M(11M - 134) + 622) - 2615) + 4212)\cos(Q) +$$
$$2(M(M(3M(3M - 37) + 538) - 1208) + 1048)\cos(2Q) + 2(451 - 6M(M(3M - 26) + 76))\cos(3Q) +$$
$$5(M - 3)(3M - 8)\cos(4Q) - 2(M - 3)\cos(5Q) - 4155M + 2904)/D_4 \quad \text{(S12)}$$

where $D_4$ is the common denominator

$$D_4 = (-2(M - 3)\cos(Q) + (M - 6)M + 10)^2(-4(M - 2)\cos(Q) +$$
$$(M - 4)M + \cos(2Q) + 7)(-2(M - 3)\cos(2Q) + (M - 6)M + 10) \quad \text{(S13)}$$

With the fourth-order correction, we find that the perturbation theory agrees very well with the numerical results as demonstrated in the main text as well as Figs. S2 and S3. However, fourth order perturbation theory for the velocity renormalization only qualitatively predicts the magic-angle transition where $\tilde{v} = 0$, but at a much larger $W$ than indicated by numerical results. It is natural to expect that this is due to the single node nature of the bandstructure at $\mathbf{M} = (\pi, \pi)$ (all of the scattering is intranode). We anticipate a better prediction of magic angle transition may be achievable only at even higher orders of perturbation theory.

Using exactly the same procedure we can consider the case of $M$ near 2, which is the SM line that divides the two TI regions with opposite sign in the quantum spin Hall effect. From a symmetry point of view it is not surprising that the $M = 2$ SM line is $W$-independent. This is indeed the case from the perturbation theory, as up to fourth order we have $\Sigma_z^{(2)} = 0$ and $\tilde{M} = M = 2$, hence there is no topological mass renormalization. That means starting from such a SM phase, quasiperiodicity is not driving it out of SM due to curvature in the phase boundary. At second order the velocity only reduces but does not go to zero and the renormalization of $\tilde{v}$ is only due to the quasiparticle residue $\Sigma_E^{(2)} = -\csc(Q/2)^2$, while $\Sigma_p^{(2)} = 0$. This can be understood as follows, at $M = 2$ the two Dirac cones are at $\mathbf{X} = (\pi, 0)$ and $\mathbf{Y} = (0, \pi)$ and being separated by $2\pi$ Manhattan distance in momentum space, second order perturbation theory will not be able to induce internode scattering. Whereas, at fourth order, the two Dirac cones can be connected by $2Q$ hops in the Brilliouin zone. Thus, only fourth order perturbation will be able to predict a vanishing velocity and a magic-angle transition. In line with this reasoning, the renormalized velocity $\tilde{v}$ up to fourth order is

$$\tilde{v} = (v + \Sigma_p^{(2)} + \Sigma_p^{(4)})/Z \quad \text{(S14)}$$

Here $\Sigma_p^{(2)}$ vanishes, and the fourth order term

$$\Sigma_p^{(4)} = \frac{W^4}{16^2}(1 + 4\cos(Q))\csc(Q/2)^4 v \quad \text{(S15)}$$

is negative only when $Q > \cos^{-1}(1/4) \approx 1.82$. Only in this regime does the perturbation theory predict a magic-angle transition. For example, at $Q = 2\pi F_{n-2}/F_n$, it predicts $\tilde{v}$ to vanish at $W_c^{(4)} = 4\sin(Q/2)(-1 - 4\cos Q)^{-1/4} \approx 3.16$. This fourth order perturbative result $W_c^{(4)}$ is an overestimate of the true critical $W_c$, and thus a more accurate prediction will require higher order perturbation theory. For smaller $Q$ such as $Q = 2\pi F_{n-3}/F_n$, the velocity can never reach 0 at fourth order in perturbation theory. Hence, the magic-angle transition is an even higher order effect than that of $Q = 2\pi F_{n-2}/F_n$. In other words, the reduction of the value of $Q$ requires higher order in perturbation (more $Q$ "hops") to capture internode scattering.

## V. FINITE SIZE EFFECTS

### A. Density of states and conductivity

In this section we discuss the finite size effects in the KPM calculations of the density of states and the conductivity. The numerical calculation puts the model on a lattice with a linear size $L$, while the truncation of the polynomial

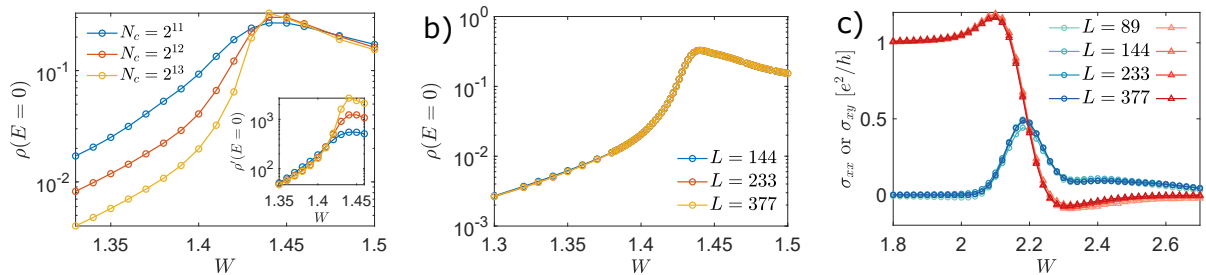

FIG. S7. **Finite size dependence of the density of states and conductivity.** (a) The $N_c$ and (b) $L$ dependence of $\rho(E = 0)$ for $M = 2$ as a function of $W$. As the expansion order $N_c$ is increased, the rise of $\rho(E = 0, W)$ is much more steep right before the transition. The inset shows $\rho'(E = 0)$, computed as $\rho' = N_c\rho$, in the regime this is $N_c$ independent we can extract the renormalized velocity $\tilde{v}$ from this scaling. The $L$ dependence is very weak for $N_c = 2^{13}$. (c) The $L$ dependence of $\sigma_{xy}$ and $\sigma_{xx}$ as we fix $N_c = 2^{10}$ for the cut $M = 3.3$. These results demonstrate that we are close to converged in $L$ and the largest finite size effect in the data stems from the finite KPM expansion order.

expansion at expansion order $N_c$ in the KPM controls the energy resolution. As $N_c$ and $L$ increase (attempting to approach the thermodynamic limit), the observables we calculate using the KPM (i.e., the density of state $\rho(E)$, the resistivity $\rho_{xx}$, the longitudinal and Hall conductivities $\sigma_{xx}$ and $\sigma_{xy}$) all become sharper, allowing us to accurately determine the critical points. In addition, we average over more than 200 samples with random twisted boundary conditions, allowing us to suppress finite size effect rooted from finite $L$ and to avoid emphasizing results from any specific choice of phase in quasiperiodicity.

As an example, we show in detail the $N_c$ and $L$ dependence near the SM-to-CM transition at $M = 2$, see Fig. S7(a,b). For fixed $L = 233$ and varying $N_c$, we see that the behavior of the density of states sharpens near the transition, demonstrating that the transition occurs between $W = 1.46$ and $W = 1.48$. Also we see that at least for density of states at zero energy, taking $L = 144$ is adequate for $N_c$ up to $2^{13}$, as the $\rho(E = 0, W)$ curve does not change at all as we vary $L$ above 144. To determine the renormalized velocity $\tilde{v}$ from the density of states we use the scaling of the SM $\rho(E) \sim \tilde{v}^{-2}|E|$ to obtain the scaling with the KPM expansion order $\rho(0) \sim \tilde{v}^{-2}(1/N_c)$. Here, we have used the fact that the infared low-energy scale ($\delta E$) induced by the finite KPM expansion order is related to $N_c$ via $\delta E = \pi D/N_c$ where $D$ is the total bandwidth of the Hamiltonian. As shown in the inset of Fig. S7(a), we find excellent data collapse of $N_c\rho(0)$ that we use to extract $\tilde{v}$ in the SM phase.

In the main text, we have shown examples of the $N_c$ dependence of the conductivity and the resistivity, where a larger cutoff leads to a steeper change in the $\rho_{xx}(W)$ curve near the critical $W$, and a longer range of $W$ where $\sigma_{xy}$ is quantized. In Fig. S7(c) we show the $L$ dependence of the longitudinal and Hall conductivities near $W_c$ as we vary $L$ but fix $N_c$. In this data we see that for $L = 89$ there is a slight variation from the other values of $L$, but above $L = 144$ the $\sigma_{xx}(W)$ and $\sigma_{xy}(W)$ are not changing at all for the different $L$'s. Again this justifies the use of $L = 144$ for $N_c$ up to $2^{10}$ to calculate conductivity as in the main text.

### B. Inverse Participation Ratio

The inverse participation ratios (IPR) reflects the structure of the wavefunction in a particular basis and is commonly used to study Anderson localization transitions. When the IPR is independent from system size, the system is localized; when the IPR scales as $1/L^d$ where $L^d$ is the volume of the $d$-dimensional lattice, the system is extended. When the IPR vanishes with a power law less then $d$, the wavefunction is critical. To consistently assess the behavior of the IPR, we calculate the IPR (in both real and momentum space bases) for each combination of parameters at $L = 55$, 89, 144 and 233. Then we fit the log of the IPR $\log(\mathcal{I})$ vs $\log L$, and extract the slope, which is the power law. We demonstrate some examples of such fitting in Fig. S8. The results of the IPR scaling shown in the main text and in Fig. S11 are all computed in this way.

## VI. EXTRACTING THE SPECTRAL GAP

The size of the insulating gap corresponds to the topological mass $\tilde{M}$; and for the topological insulator to metal phase transition the power-law scaling of the gap size allows us to extract critical exponents at the TI to CM transition as described in the main text. Here we elaborate on the details for extracting spectral gap.

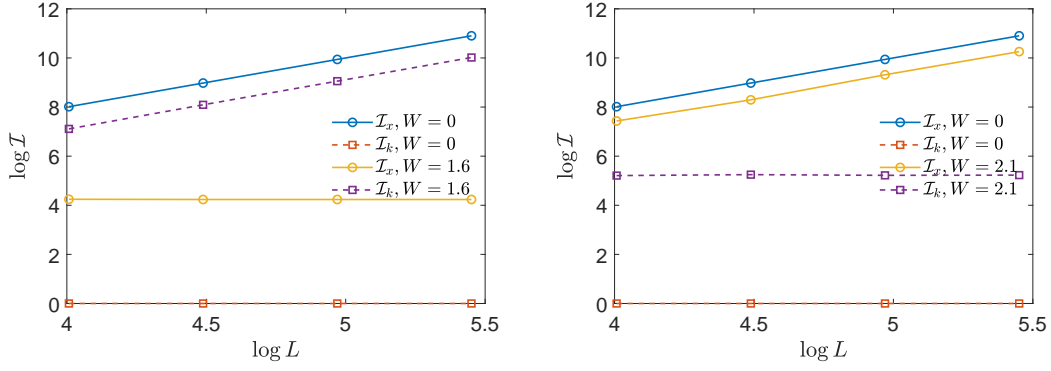

FIG. S8. **L dependence of the IPR** Here we demonstrate two examples of how we determine $\gamma_\alpha$ for IPR data in the basis $\alpha = x, k$, where $I_\alpha \sim L^{-\gamma_\alpha}$, for $M = 2.7$ (a) and $M = 3.3$ (b). We take a linear fit for $\log I_k$ or $\log I_x$ over $\log L$, then the slope of the fit estimates $\gamma$.

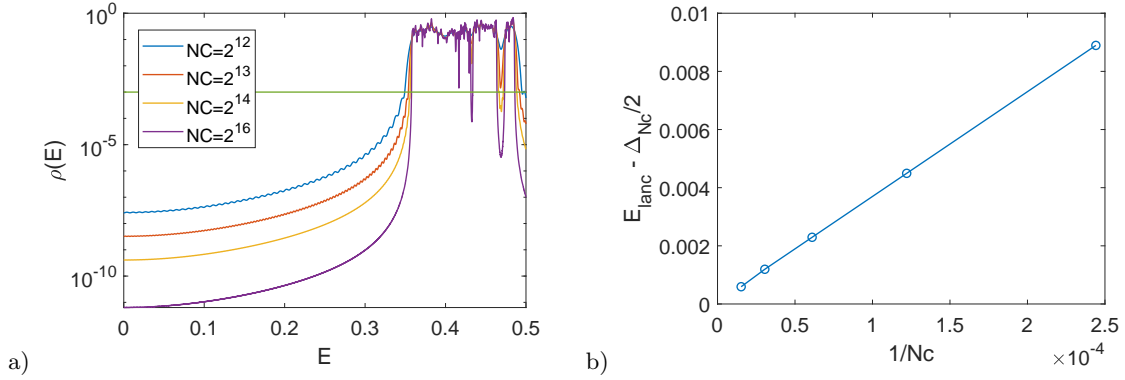

a)                                                            b)

FIG. S9. **Extracting gap from KPM.** Simply choosing a threshold $\varepsilon$ makes controlled approximation of gap size in KPM, and matches Lanczos result at large $N_c$ limit. a) Density of state $\rho(E)$ in log scale. The green line marks the threshold $\varepsilon = 0.001$ used for determining $\Delta$ using KPM. The crossing point between the green line and $\rho(E)$ curve marks the upper end of the gap; while the lower end is determined similarly. b) The distance between half of the gap size determined by KPM with different cutoff ($\Delta_{N_c}/2$) to the lowest energy state determined by Lanczos ($E_{\text{lanc}}$) as a function of $1/N_c$. At large $N_c$ limit, the difference approaches zero. For $N_c = 8192$ as an example, the KPM calculated $\Delta$ is accurate up to $1e - 3$.

The size of the insulating gap centered at $E = 0$ is most efficiently calculated by extracting from DOS data in general. We first use the integral of density of state $n(E_1, E_2) = \int_{E_1}^{E_2} \rho(E) dE$ to find a positive energy $E_{\text{max}}$ such that $n(0, E_{\text{max}})$ is smaller than $1/2L^2$. Then the range inside $(-E_{\text{max}}, E_{\text{max}})$ with $\rho(E)$ below the chosen threshold $\varepsilon$ is the gap. The gap size $\Delta$ is the length of the range. Fig. S9 (a) illustrates how the gap is extracted.

In general, KPM involves a broadening of $N_c$, and hence the choice of threshold would depend on $N_c$. However, as long as $1 >> \varepsilon >> 1/N_c$, the numerical result of $\Delta$ from KPM with $N_c$ provide an estimation approximately good up to $1/N_c$. We can also use Lanczos method by setting shift-invert factor $\sigma$ to be KPM estimated gap edge to obtain the accurate energy of lowest energy state above $E = 0$, which is half of gap size. Fig. S9(b) demonstrate that the difference between KPM estimated gap size and the accurate Lanczos result. As $N_c$ approaches infinity, the differnce vanishes.

In fact, for $W$ very close to the TI-CM transition, Lanczos method become more efficient than KPM to accurately determine $\Delta$. Using shift-and-invert Lanczos with shift factor $\sigma = 0$, we can find the minimum of the lowest energy band. Here, we demonstrate a few examples of the gap size extracted using the Lanczos method with various $L$ in supplement to the gap size extracted in the main text using KPM. As shown in Fig. S10(a-c) the result is essentially $L$ independent.

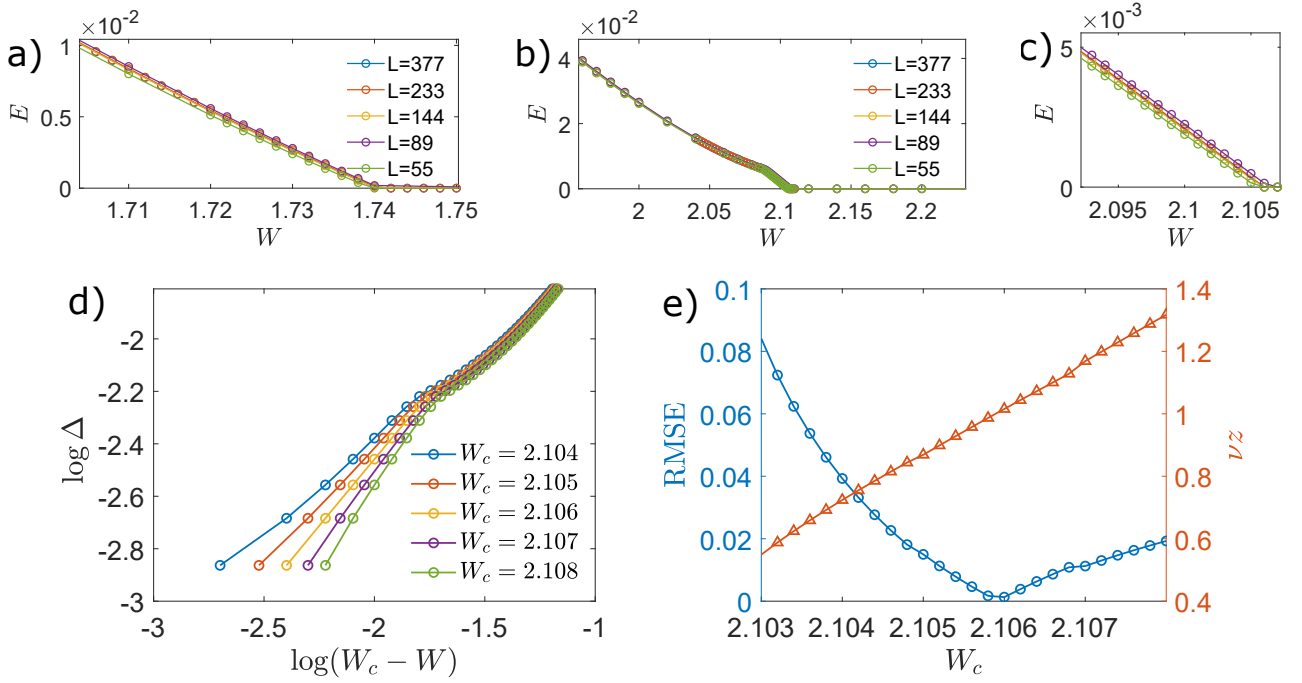

FIG. S10. **Vanishing of the spectral gap.** The gap size as a function of $W$ for $M = 2.4$ (a) and $M = 3.0$ (b) with various system sizes $L$. (c) is a more zoomed in view of the $M = 3.0$ cut near the transition. In the second row, we show an example of how the combination of critical exponents $\nu z$ is extracted from the spectral gap data. For a range of choices of $W_c$, we fit $\log \Delta$ against $\log(W_c - W)$ in the range when $W_c - W$ is under 0.015 and $\Delta > 0.001$ with a straight line. This data is shown in (d). Then we find the point where the root mean square error (RMSE) as shown in (e) of the linear fit is smallest as our best estimation of $W_c$, where the slope is then $\nu z$. For these results, our best estimate of $\nu z = 1.0 \pm 0.1$, with $W_c = 2.106 \pm 0.001$.

### A. Critical Exponents at the topological insulator to metal phase transition

The critical exponents at the TI-to-CM phase transition can be extracted from the gap size, as quoted in the main text. Here we demonstrate this process in more detail. The $\Delta$ is calculated using Lanczos here, as we focus on a very small range of parameters, making Lanczos more efficient than KPM. We first estimate the critical quasiperiodic strength $W_c$ from the density of states data. Near the estimated $W_c$, we consider a range of $W_c$ and fit $\log(\Delta)$ over $\log(W_c - W)$, see Fig. S10(d). Then we identify the range of $W$ that $\log \Delta$ is linear to $\log(W_c - W)$ and use least square fit. The best $W_c$ is picked according to the goodness of linear fit, here quantified with root mean square error of the fit, see Fig. S10(e). The slope of the best fit is then $\nu z$, where $\Delta \sim (W_c - W)^{\nu z}$. For some cases this critical exponent is difficult to determine accurately because of the very fine phase diagram structure. For a few fixed $M$ cuts shown in the main text, including $M = 2.4$, $M = 3.0$, $M = 3.3$ and $M = 3.8$, we find $\nu z$ near 1.0. To be precise, for $M = 2.4$ we have $\nu z = 1.06 \pm 0.1$; for $M = 3.0$, $\nu z = 1.00 \pm 0.1$; for $M = 3.3$, $\nu z = 0.95 \pm 0.2$; for $M = 3.8$, $\nu z = 1.13 \pm 0.15$

### B. Gapsize and IPR

Another interesting feature is the boundary where our perturbation theory works to determine the gap size. The perturbation theory expression of the gap size gives a qualitatively correct prediction of the dependency of $\Delta$ on $W$, but the trend has a turning point at some finite $W$ after which it no longer follows the perturbation theory. This is when the lowest band is no longer clearly the topological band, but mixes with other minibands nearby in energy due to the quasiperiodic potential. We show that the IPR of the lowest energy state changes dramatically at the same $W$ where the gap size turns down as shown in Fig. S11.

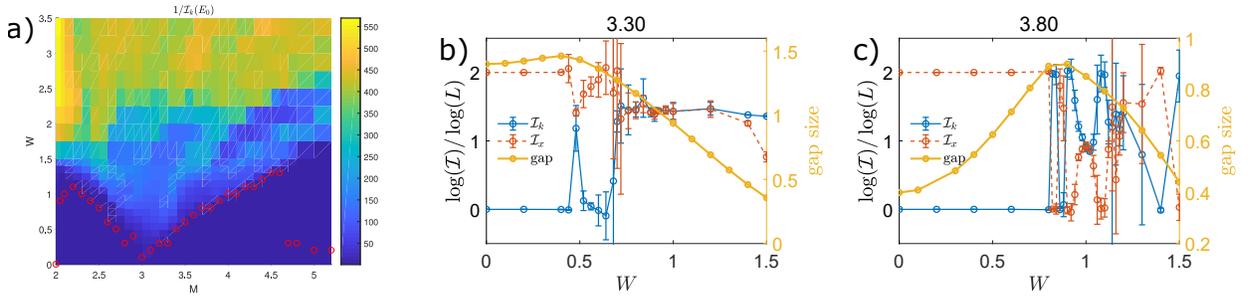

FIG. S11. **Properties of the IPR.** (a) Phase diagram of the momentum space IPR of the lowest eigenstates. The red circles mark where the gap size $\Delta(W)$ changes its trend from increasing to decreasing as determined by the location of the maximum in $\Delta'(W)$. The cuts $M = 3.3$ (b) and $M = 3.8$ (c) show the non-trivial $L$ dependence of the IPR in both real and momentum space start to dramatically change when $\Delta(W)$ begins to turn downward.

## VII. DISPERSION RELATION, EFFECTIVE MASS AND BERRY CURVATURE

In this section, we discuss our calculation of eigenstates and wavefunction, and hence dispersion relations and Berry curvature. The twisted boundary condition we implemented, $t_\mu \rightarrow t_\mu e^{i\theta_\mu/L}$, effectively shift momentum $\mathbf{k}$ by $\boldsymbol{\theta}/L$ . Without a quasiperiodic potential, this corresponds to a trivial folding of the original energy disperion $E(\mathbf{k})$ such that $E_n(\boldsymbol{\theta}) = E(2\pi n/L + \boldsymbol{\theta}/L)$ for band $n$ in the Brillouin zone defined by the supercell of size $L \times L$. In other words, $\boldsymbol{\theta}/L$ is the momentum in the mini-Brillouin zone of the superlattice made of $L \times L$ supercells.

To calculate eigenvalues and eigenstates in the interior of the spectrum, we use shift-invert Lanczos method with a shift factor $\sigma$. The spectrum is quite dense, so we can only accurately calculate eigenstates near a hard gap in order to have a definite band index for each state as we go to a large system size. Hence, $\sigma$ is set to be the energy of the desired band edge next to the gap, which can be accurately located using DOS calculated with the KPM. Such a shift factor allows fast convergence to the first $n$ eigenpairs with definite band indices, as long as the $E_i - \sigma$ for all $i \leq n$ is less than the gap size. The computational complexities in both memory and time are dominated by the Cholesky decomposition step in shift-invert Lanczos; whereas the iterations over the Lanczos Ritz vectors to increase the accuracy of eigenvalue and eigenvector calculation are fast. Hence we can always set the error tolerance of Lanczos to be at least $1/100$ of the bandwidths of the folded bands.

### A. Effective mass

One additional measure of band flatness to consider is the effective mass of the lowest band. We define an effective mass $m^*$ by the expansion of the energy dispersion about its minimum $E(\mathbf{q}) = E(0) + \frac{1}{2m^*}\mathbf{q}^2 + \cdots$. Numerically, $m^*$ can be obtained through a quadratic fitting of $E_n(\boldsymbol{\theta})$ near $\boldsymbol{\theta} = \mathbf{0}$, where the $n$ indexes the first, folded band above $E = 0$ (Fig. S12(a), for example). The effective mass can also be obtained, perturbatively, from the pole of the Green function $E_{\text{eff}}(\mathbf{q}) = \pm(\tilde{M} + \mathbf{q}^2/2m^*)$ where $m^* = \tilde{M}/(2\tilde{v}^2)$. At fourth-order, $m^* \sim 10^5(1/t)$ at $W = 3t$, indicating that the QP is flattening the topological bands; our numerics (the blue solid line in Fig. S12(b)) show this effect is even more drastic [Fig. S12(b)].

### B. Edge states

For finite size calculation, (twisted) periodic boundary conditions eliminate all edge contribution and keep only the bulk. With open boundary conditions, the edge states can be observed, but we no longer have access to any $\boldsymbol{\theta} \neq \mathbf{0}$. Here, we use twisted boundary conditions only in the $x$-direction but open boundary conditions along the $y$-direction, so that we can see the dispersion of the edge states as we vary $\theta_x$, as clearly demonstrated in Fig. S12(c).

### C. Berry curvature of folded bands

To clearly characterize the topology of the folded bands, we calculate the Berry curvature directly by replacing momentum with the corresponding twist in the boundary condition. We dived the minizone of size $(2\pi/L)^2$ into a

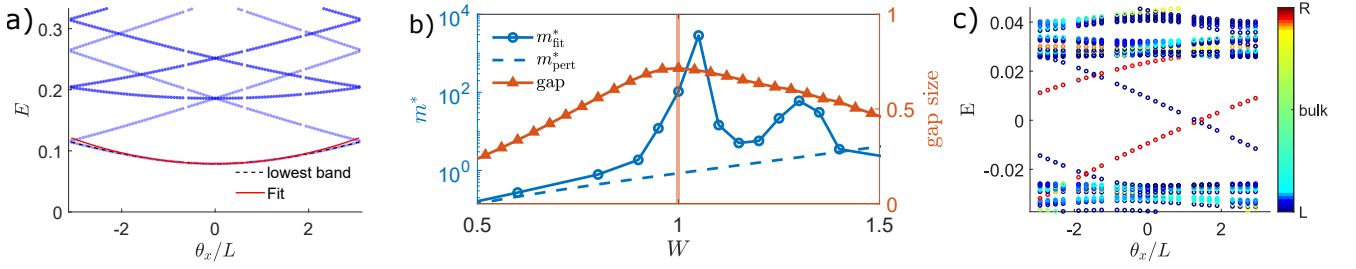

FIG. S12. **Twist Dispersions**. (a) Using the twist dispersion to obtain the effective mass $m^*$. The red curve is the quadratic fitting result to estimate $m^*$. The figure shows an example for $M = 4.0$, $W = 0.4$. (b) The effective mass obtained from fitting twist dispersion ($m^*_{\text{fit}}$) and from perturbation theory ($m^*_{\text{pert}}$), compared with gap size. The vertical line marks the $W$ where we scrutinize flat topological bands in the main text. (c) Twist dispersion with open boundary conditions in the $y$ direction and twisted boundary conditions in the $x$ direction. The color corresponds to the location of the eigenstates along the $y$ axis. The red and dark blue states in the bulk gap are the edge states.

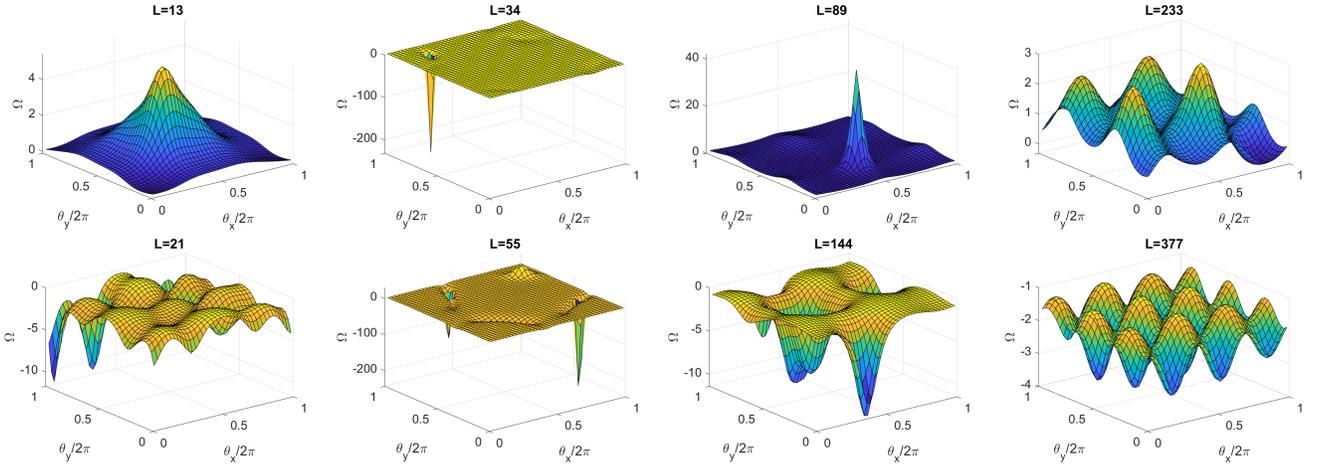

FIG. S13. **Berry curvatures**. The Berry curvature of of the first band above the hard gap near $E = -0.5$. The samples shown are at $W = 1.01541$, $M = 4$ and $L = 377$, i.e. at its peak flatness (see Fig. 3(b) of the main text). The first row are system sizes in the sequence of $L = F_n$ with odd $n$, and the second row for even $n$. For $L = 55$ and $L = 89$, Berry curvature have clear peaks; while larger $L$'s see flatter Berry curvature.

grid of size $N_\theta \times N_\theta$ and evaluate $|\psi_n(\boldsymbol{\theta})\rangle$ with $N_\theta = 40$, where $n$ is the band index starting from the band edge near a large gap. We follow Ref.[S7] to obtain a manifestly gauge invariant calculation that always sum to integer valued Chern number. Note that the method requires the folded band to be gapped, so for example, at small $W$ the calculation will fail. In addition, using Lanczos to compute the eigenstates leaves a phase ambiguity that may change for each calculation. When calculating the Berry curvature on any plaquette, each of the four $|\psi_n(\boldsymbol{\theta})\rangle$ surrounding the plaquette is used twice and it is important to use the same $|\psi_n(\boldsymbol{\theta})\rangle$ to be consistent. At the same time, calculations for different plaquettes are mutually independent, allowing for a straightforwardly parallelized calculation. This can be done either by saving $|\psi_n(\boldsymbol{\theta})\rangle$, or carefully setting the initial vector and the random seed for Lanczos.

Some examples of the calculated Berry curvature are shown in Fig. S13. For each case, the Berry curvature in a given band sums up to the Chern number, which is used for determining the topology of the band structure in main text. To characterize the fluctuations of Berry curvature, the normalized standard deviation of Berry curvature, defined as

$$\frac{\text{s.d.}(\Omega_n)}{\bar{\Omega}_n} \equiv \frac{\sqrt{\sum_{\boldsymbol{\theta}} \Omega_n(\boldsymbol{\theta})^2 - \left(\sum_{\boldsymbol{\theta}} \Omega_n(\boldsymbol{\theta})\right)^2}}{\sum_{\boldsymbol{\theta}} \Omega_n(\boldsymbol{\theta})} \tag{S16}$$

is used.

---



\end{document}